# AI Models Still Lag Behind Traditional Numerical Models in Predicting Sudden-Turning Typhoons

Daosheng Xu[a,#], Zebin Lu[b,#], Jeremy Cheuk-Hin Leung[c], Dingchi Zhao[c], Yi Li[c,*], Yang Shi[d], Bin Chen[e], Gaozhen Nie[f], Naigeng Wu[d], Xiangjun Tian[g], Yi Yang[e], Shaoqing Zhang[b], Banglin Zhang[c,e,*]

**Affiliations**

[a] Guangzhou Institute of Tropical and Marine Meteorology/Guangdong Provincial Key Laboratory of Regional Numerical Weather Prediction, China Meteorological Administration, Guangzhou, China

[b] Key Laboratory of Physical Oceanography, Ministry of Education/Institute for Advanced Ocean Study/Frontiers Science Center for Deep Ocean Multispheres and Earth System (DOMES), College of Oceanic and Atmospheric Sciences, Ocean University of China, Qingdao, China.

[c] College of Meteorology and Oceanography, National University of Defense Technology, Changsha, China

[d] Guangdong Meteorological Observatory, Guangzhou, China

[e] College of Atmospheric Science, Lanzhou University, Lanzhou, China

[f] National Meteorological Centre, China Meteorological Administration, Beijing, China

[g] State Key Laboratory of Tibetan Plateau Earth System, Resources and Environment (TPESRE), Institute of Tibetan Plateau Research, Chinese Academy of Sciences, Beijing, China

[#] These authors contributed equally to this work.

*Correspondence to: zhangbanglin24@nudt.edu.cn (Banglin Zhang), *liyiqxxy@163.com* (Yi Li)

Given the interpretability, accuracy, and stability of numerical weather prediction (NWP) models, current operational weather forecasting relies heavily on the NWP approach[1]. In the past two years, the rapid development of Artificial Intelligence (AI) has provided an alternative solution for medium-range (1~10 days) weather forecasting. Bi et al. [2] (hereafter Bi23) introduced the first AI-based weather prediction (AIWP) model in China, named Pangu-Weather, which offers fast prediction without compromising accuracy. In their work, Bi23 made notable claims regarding its effectiveness in extreme weather predictions. However, this claim lacks persuasiveness because the extreme nature of the two tropical cyclones (TCs) examples presented in Bi23, namely Typhoon Kong-rey and Typhoon Yutu, stems primarily from their intensities rather than their moving paths [3]. Their claim may mislead into another meaning which is that Pangu-Weather works well in predicting unusual typhoon paths, which was not explicitly analyzed. Here, we reassess Pangu-Weather's ability to predict extreme TC trajectories from 2020–2024. Results reveal that while Pangu-Weather overall outperforms NWP models in predicting tropical cyclone (TC) tracks, it falls short in accurately predicting the rarely observed sudden-turning tracks, such as Typhoon Khanun in 2023. We argue that current AIWP models still lag behind traditional NWP models in predicting such rare extreme events in medium-range forecasts.

To gain a deeper understanding of Pangu-Weather's ability to forecast extreme TC cases relative to its counterparts, we evaluate its performance in predicting Northwest Pacific TCs from 2020–2024. We divide all TCs into three categories, according to their moving trajectory: (1) ordinary TCs, (2) sudden-turning TCs, and (3) looping TCs. The latter two types belong to unusual TC trajectories, which occur less frequently and are often more difficult for forecasters to accurately predict their tracks. Among the 104 TC cases analyzed in this study, there are 14 sudden-turning TCs, 7 looping TCs, and the rest (83) are categorized as ordinary TCs. And, in the following discussion, four types of prediction approaches are compared: (1) Pangu-Weather, including Pangu-ERA5, Pangu-ECMWF, and Pangu-NCEP; (2) global NWP, including ECMWF-IFS and NCEP-GFS; (3) regional NWP, i.e., CMA-TRAMS-L125 which is driven by ECMWF-IFS; and (4) human forecasters, which are the official forecasts issued by JTWC, JMA, and CMA (see Supplementary Texts S1 and S2 for definitions).

Overall, Pangu-Weather performs exceptionally well in predicting TCs' trajectories, especially in 2023 (Fig. S1). Pangu-ERA5 exhibited greater accuracy than ECMWF-IFS, which is consistent with Bi23's conclusion [2]. Although the accuracies slightly drop when Pangu-Weather is driven by real-time ECMWF-IFS analysis [4], Pangu-ECMWF still outperforms ECMWF-IFS for most TCs (Figs. S2 and S3). Despite the overall better performance, Pangu-Weather had larger errors in predicting uncommon TC trajectories, such as Severe Typhoon Khanun (202306). It made two sharp turns within 5 days when it passed through the Ryukyu Islands (Fig. 1a) and caused significant damage in surrounding countries. Despite its weaker intensity compared to Kong-rey and Yutu, Khanun's exceptional path rendered it an extreme event, posing challenges for accurately predicting its movements and landfall position. In this case, Pangu-ERA5 and Pangu-ECMWF exhibit average 24–120 h track forecast errors 7.0% and 18.8% greater than those of ECMWF-IFS, respectively (Fig. 1e).

Digging deeper, Pangu-Weather demonstrated smaller average biases and uncertainties compared to human forecasters, but greater than NWP models, for the 24-to-120-hour track forecasts of Khanun (Figs. 1b–d). Furthermore, compared with both human forecasters and NWP models, Pangu-Weather exhibited a steeper decline in prediction skill as forecast lead time increased (Fig. 1f, Supplementary Text S3 and Fig. S4). Consequently, the difference between Pangu-ECMWF and human forecasters becomes indistinguishable for 5-day forecasts (Fig. 1e). These findings indicate Pangu-Weather's limitations in predicting rarely-occurring extreme TC cases. Although Pangu-Weather demonstrates good prediction skills in global-scale forecasts [2], it only catches the performance of NWP models for short-term forecasts of 1-2 days in extreme cases like Typhoon Khanun, but not for longer lead times.

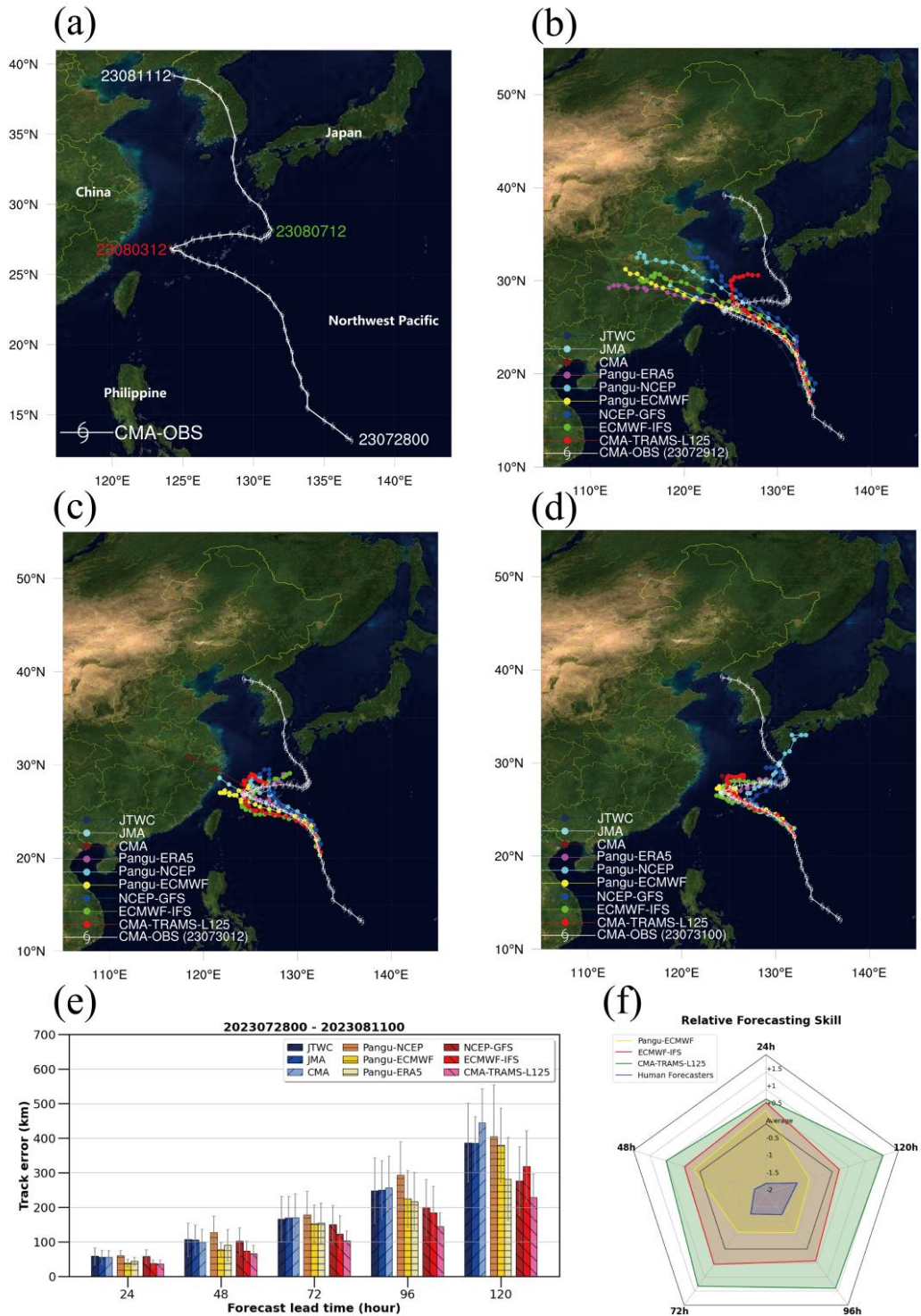

**Fig. 1.** Performances of NWP models, AIWP models, and human forecasters in predicting Typhoon Khanun's trajectory. (a) Typhoon Khanun's trajectory during its lifetime, according to the CMA observation. (b–d) 168-hour prediction of Typhoon Khanun's track by JTWC, JMA, CMA, Pangu-NCEP, Pangu-ECMWF, Pangu-ERA5, NCEP-GFS, ECMWF-IFS, and CMA-TRAMS-L125 at (b) 1200UTC 29th July, (c) 1200UTC 30th July, (d) 0000UTC 31st July, respectively. (e) Average track forecast errors (bars, unit: km) and their standard deviation (error bars, unit: km) of human forecasters (issued by JTWC, JMA, and CMA), AIWP models (Pangu-NCEP, Pangu-ECMWF, and Pangu-ERA5), global NWP models (NCEP-GFS and ECMWF-IFS), and a regional NWP model (CMA-TRAMS-L125). (f) Relative forecasting skills (Supplementary Text S2) of ECMWF-IFS, Pangu-ECMWF, CMA-TRAMS-L125, and human forecasters (average of JTWC,

Typhoon Khanun is not the only example of sudden-turning TCs. Among the 104 TC cases during 2020–2024, there are 14 sudden-turning TCs (Fig. 2b). We find that ECMWF-IFS overall outperforms Pangu-ECMWF for these sudden-turning cases. The mean distance error of Pangu-ECMWF ranges from 40.9–400.2 km for 6–120 forecast hours, which are on average 9.4% greater than those of ECMWF-IFS (ranging from 30.1–354.0 km) (Fig. 2e). This further confirms the conclusion that Pangu-Weather performs worse, compared with NWP models, in capturing the extreme nature of these unusual TC paths.

On the other hand, it is also important to acknowledge the advantages of Pangu-Weather. Our results indicate that Pangu-ECMWF overall outperforms ECMWF-IFS in predicting ordinary TCs (Fig. 2a). The mean distance error of Pangu-ECMWF (ranging from 49.9–244.9 km) is on average 13.5% smaller than those of ECMWF-IFS (ranging from 48.8–277.2 km) (Fig. 2d). Moreover, Pangu-ECMWF also performs well in predicting looping TCs (Fig. 2c), which are also considered anomalous cases. The mean distance error of Pangu-ECMWF (ranging from 52.8–240.5 km) is on average 32.9% smaller than those of ECMWF-IFS (ranging from 40.8–403.2 km) (Fig. 2f). Analyses based on other two AIWP models, Fengwu [5] and FuXi [6], yield consistent results.

The above results suggest that AIWP models underperform NWP models in a particular type of TC, that is sudden-turning typhoons. A key question is, what is the key factor limiting AIWP models' forecast skills in predicting this particular type of extreme TC trajectory?

We find that TCs that are better predicted by ECMWF-IFS predominantly exhibit usual trajectories, following the large-scale climatological steering flow of the western Pacific subtropical high (WPSH). They are associated with background southeasterly winds and a stronger WPSH (Fig. 2g). Under the influence of a stronger WPSH, TCs' moving trajectories are more likely to follow the background steering flow. In contrast, for TCs that are better predicted by Pangu-ECMWF, the guiding airflow is mainly southwesterly winds, and the WPSH is weaker. Under the background of a weaker WPSH, TCs' trajectories are often the result of the interaction between the background steering flow and the TC itself, and result in anomalous paths. In this case, the prediction accuracies of the fine structural characteristics of the TC core region become important (Supplementary Text S4, Figs. S5–S6).

Since AIWP models, such as Pangu-Weather, are more capable of capturing the evolution patterns of large-scale circulations (e.g., WPSH), they tend to give better prediction results for TCs when the WPSH is strong. On the contrary, AIWP models are not good at "simulating" TC structures and fail to reproduce the realistic physics of TC systems. Thus, AIWP models tend to give worse prediction results for the fine structural characteristics of the TC core region when the WPSH is relatively weak.

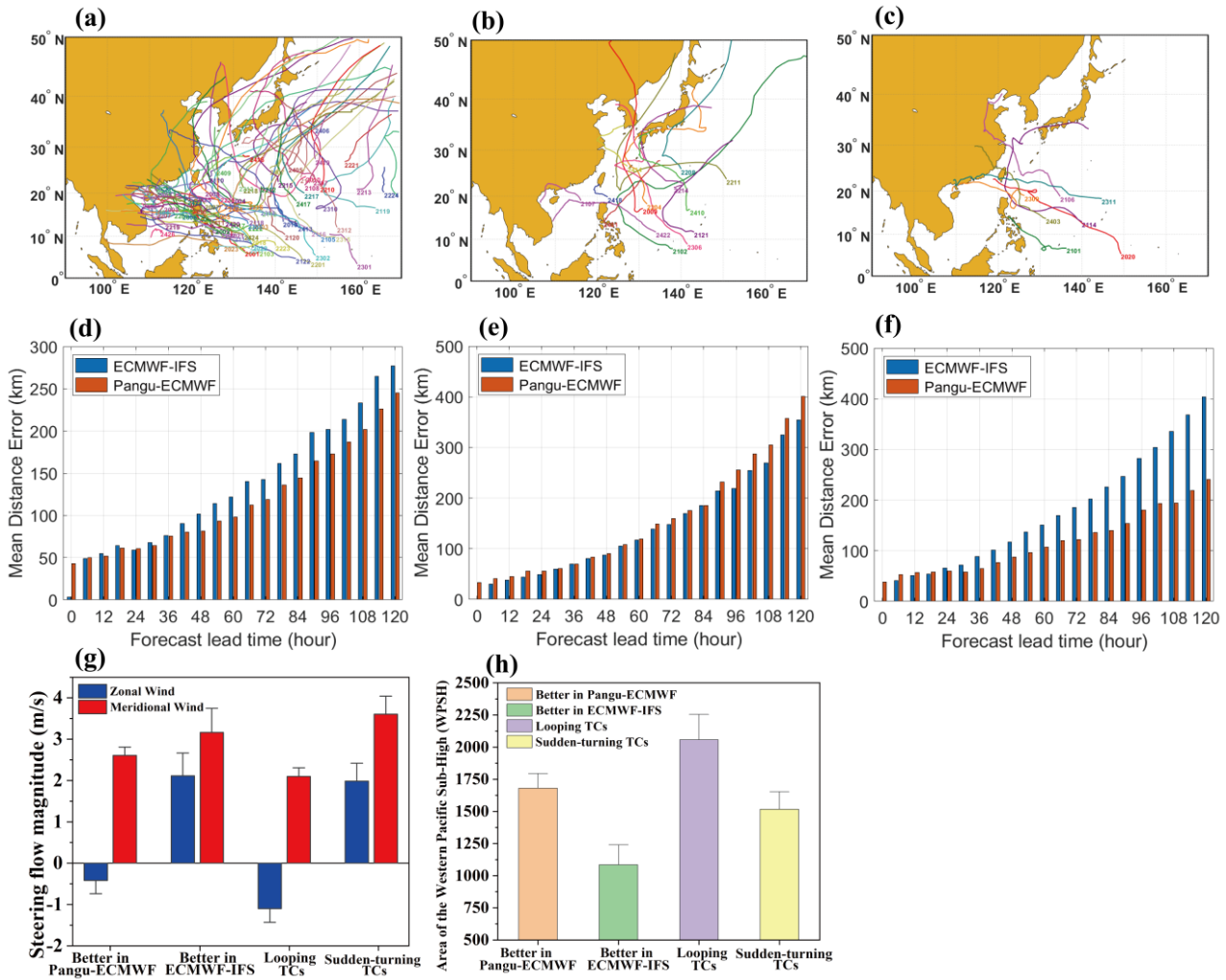

**Fig. 2.** (a–c) Trajectories of (a) ordinary, (b) sudden-turning, and (c) looping TCs from 2020 to 2024. (d–f) Same as (a–c) but for the average error (unit: km) of TC tracks predicted by the ECMWF-IFS (blue) and Pangu-ECMWF (red). (g) Composited 200–850 hPa zonal (wind) and meridional (red) wind (unit: m/s) of different groups of TCs. Note the composite values are calculated by the averaged wind field within 500 km from the centers of TCs. (h) Composited area (number of grids with geopotential exceeding 5880 gpm) of WPSH for different groups of TCs. Note the composite values are calculated by the WPSH within 20°×20° from the centers of TCs. Results are based on the ERA5 data.

Unlike NWP models which are designed to solve a set of partial differential equations governing the atmosphere, AIWP models are data-driven models trained to make predictions by learning historical weather patterns. This notably different approach may lead to several potential downsides limiting AIWP models' ability to predict extreme events. For example, the fully data-driven training approach makes AIWP models more adept at capturing typical weather patterns, but less effective at predicting extreme cases. The limited samples of extreme events make AIWP models hard to predict patterns that rarely occurred in the past. In addition, Pangu-Weather exhibits apparent error in predicting the geostrophic relationship between geopotential and wind [7] (Fig. S8), as well as Khanun's strong updraft and downdraft (Fig. S9) [7,8], although it well captures the TC warm core structure and its quasi-hydrostatic approximation [9] (Fig. S7). This evidence suggests that AIWP models have the potential to generate predictions that may be unrealistic and not inherently bound by physical laws [10] (see Supplementary Text S5). Besides, the use of global root-mean-square error as the model cost function tends to penalize predictions of extreme values, hence generating relatively smooth forecast outputs, which is also evident in Fig. S7d and the underestimation of TC intensities [2]. In addition, the above issue could also smooth out atmospheric systems surrounding a TC (Fig. S9d) and subsequently influence the prediction of TC moving speed and direction.

Our analyses indicate that a major issue affecting AIWP models' prediction skills on extreme TC trajectories is their limited ability to forecast the fine structural characteristics of TCs, which lowers the ability of AIWP models to reasonably describe the interaction between TCs and large-scale steering flow when the WPSH is weak, thereby affecting the accuracy of sudden-turn predictions in TC trajectories. We believe that addressing this issue should be approached in two directions: (1) creating higher-resolution reanalysis datasets and training high-resolution AI models based on these datasets; (2) recognizing that traditional NWP models have advantages in forecasting extreme events, such as sudden turns in TC trajectories. Therefore, it is necessary to integrate more physical constraints into AIWP models, to mitigate the problem of excessive smoothing in the prediction of certain extreme events, a limitation often encountered in pure AI models with restricted sample sizes.

In conclusion, this brief commentary serves as a supplementary evaluation of Pangu-Weather's ability to predict extreme events. While Pangu-Weather exhibits higher overall prediction skills in TC track predictions, it still lags behind ECMWF-IFS in predicting sudden-turning TCs, compared to ECMWF-IFS. Especially for extreme cases like Khanun, Pangu-Weather achieves high accuracy in 1-to-2-day forecasts, but not for longer lead times. This highlights the necessity of improving AIWP models' medium-range forecasting skills under rare weather conditions. The AI-based Pangu-Weather model has already demonstrated advantages over human forecasters, demonstrating the power of AIWP models in extracting natural patterns from historical data. Moving forward, unraveling small sample size problems and incorporating physical constraints into the model training process could enhance AIWP models' ability to predict abnormal weather phenomena, to cope with the increasing risks of extreme weather caused by climate change.

## Data availability

All the data analyzed in this article are available from the corresponding author upon reasonable request.

## Acknowledgments

This study is supported by the Independent Innovation Science Fund of National University of Defense Technology (Grant Nos. 22-ZZCX-081) and the Guangdong Province Introduction of Innovative R&D Team Project China (2019ZT08G669). We would like to thank Dr. Dehui Chen from CMA Earth System Modeling and Prediction Center, who asked for us to evaluate CMA-TRAMS-125L's ability to predict Khanun's zigzagging track. This has in turn led us to further explore the comparative performance of AIWP models against traditional NWP models and forecasters.

## Competing interests

The authors declare that they have no known competing financial interests or personal relationships that could have appeared to influence the work reported in this paper.

## Author contributions

The authors confirm contribution to the paper as follows: Conceptualization: Banglin Zhang; Formal analysis: Daosheng Xu, Zebin Lu, Jeremy Cheuk-Hin Leung; Writing—original draft: Jeremy Cheuk-Hin Leung; Visualization: Daosheng Xu, Zebin Lu, Jeremy Cheuk-Hin Leung, Dingchi Zhao, Yang Shi; Data curation: Daosheng Xu, Zebin Lu, Yang Shi, Gaozhen Nie, Naigeng Wu; Writing—review and editing: Banglin Zhang, Zebin Lu, Jeremy Cheuk-Hin Leung, Dingchi Zhao, Yi Li, Bin Chen, Shaoqing Zhang; Methodology: Banglin Zhang, Daosheng Xu, Zebin Lu. Jeremy Cheuk-Hin Leung; Funding acquisition: Banglin Zhang, Yi Li, Bin Chen, Shaoqing Zhang; Supervision: Banglin Zhang, Yi Li. All authors reviewed the results and approved the final version of the manuscript.

**Figure captions**

**Fig. 1.** Performances of NWP models, AIWP models, and human forecasters in predicting Typhoon Khanun's trajectory. (a) Typhoon Khanun's trajectory during its lifetime, according to the CMA observation. (b–d) 168-hour prediction of Typhoon Khanun's track by JTWC, JMA, CMA, Pangu-NCEP, Pangu-ECMWF, Pangu-ERA5, NCEP-GFS, ECMWF-IFS, and CMA-TRAMS-L125 at (b) 1200UTC 29th July, (c) 1200UTC 30th July, (d) 0000UTC 31st July, respectively. (e) Average track forecast errors (bars, unit: km) and their standard deviation (error bars, unit: km) of human forecasters (issued by JTWC, JMA, and CMA), AIWP models (Pangu-NCEP, Pangu-ECMWF, and Pangu-ERA5), global NWP models (NCEP-GFS and ECMWF-IFS), and a regional NWP model (CMA-TRAMS-L125). (f) Relative forecasting skills (Supplementary Text S2) of ECMWF-IFS, Pangu-ECMWF, CMA-TRAMS-L125, and human forecasters (average of JTWC, JMA, and CMA) for each forecast lead time. The track forecast errors are calculated by comparing the predictions against the CMA best track data.

**Fig. 2.** (a–c) Trajectories of (a) ordinary, (b) sudden-turning, and (c) looping TCs from 2020 to 2024. (d–f) Same as (a–c) but for the average error (unit: km) of TC tracks predicted by the ECMWF-IFS (blue) and Pangu-ECMWF (red). (g) Composited 200–850 hPa zonal (wind) and meridional (red) wind (unit: m/s) of different groups of TCs. Note the composite values are calculated by the averaged wind field within 500 km from the centers of TCs. (h) Composited area

(number of grids with geopotential exceeding 5880 gpm) of WPSH for different groups of TCs. Note the composite values are calculated by the WPSH within 20°×20° from the centers of TCs. Results are based on the ERA5 data.

# Supplementary Information

## Contents of this file

Texts S1 to S5

Figures S1 to S9

Table S1

References

## Introduction

This supporting information provides supplementary texts, figures, and tables cited in the main text.

## Text S1. Data

*1. TC best track data*

The observed typhoon intensity and path data used for verification are released in real-time by the National Meteorological Centre typhoon network (http://typhoon.nmc.cn/web.html). In 2023, a total of 17 named TCs were recorded in the Northwest Pacific, of which 15 were included in the evaluation analyses. TC Dora (2308) was excluded because of its short duration and TC Jelawat (2317) was excluded because it did not impact East Asia.

*2. NWP forecast data*

After decades of 'quiet' evolution, the performance of numerical weather prediction (NWP) has been steadily improving thanks to advancements in forecasting algorithms and data assimilation techniques, more available observations, and enhanced computing power [1]. In this commentary, two global NWP models and one regional NWP model were evaluated. These NWP models simulate weather evolution by solving a set of atmospheric governing equations that strictly follow the laws of physics, in contrast to the solely data-driven Pangu-Weather model. All NWP models provide 6-to-168-hour prediction outputs, with a 6-hour temporal resolution.

The NWP forecast data for this study include the operational TC track forecasts of two global NWP models (ECMWF-IFS, NCEP-GFS), and a regional NWP model (CMA Tropical Regional Atmosphere Model for the South China Sea, CMA-TRAMS-125L).

We use ECMWF-IFS operational forecasts issued at the time from the high-resolution deterministic (HRES) version. The forecast data has a horizontal resolution of 0.1 × 0.1° and 17 vertical levels. The forecast data are issued twice a day at 0000 UTC and 1200 UTC, and the forecast lead time ranges from 24 h to 240 h (10 days). A detailed description of the IFS can be found at https://www.ecmwf.int/en/publications/ifs-documentation.

The NCEP-GFS operational forecast has a global prediction mission for medium-range (3–14 days) forecasts. Its output is posted to a 0.25° equally spaced in longitude/latitude with a 6-h forecast interval to 180 h, cycled four times a day, with 37 vertical standard pressure levels. The NCEP-GFS forecast data can be downloaded at https://rda.ucar.edu/datasets/ds084.1/dataaccess/#.

The CMA-TRAMS model, with a horizontal resolution of 0.09°, uses the ECMWF-IFS as both the initial and boundary conditions for operational implementation. Because of the slow data transfer speed, the ECMWF data are selected at a very low vertical resolution of 17 levels.

*3. Pangu-Weather forecast data*

The Pangu-Weather model, similar to other AI-based weather prediction (AIWP) models such as FourCastNet [2] and GraphCast [3], demonstrates the ability to generate forecast results with overall accuracies comparable to those of the leading numerical weather prediction (NWP) model, namely, the integrated forecasting system of the European Centre for Medium-Range Weather Forecasts (ECMWF-IFS). These AIWP models achieve such impressive accuracies while consuming only a fraction of the computational resources required by traditional NWP methods. The remarkable success of these lightweight, fast, and accurate AIWP models presents abundant opportunities for advancing the field of weather forecasting, which has traditionally progressed at a slower pace [4].

With the help of the Huawei Cloud Team, Pangu-Weather was successfully installed at the Guangdong Meteorological Observatory and operated with various initial data, including the European Centre for Medium-Range Weather Forecasts (ECMWF) Reanalysis v5 (ERA5) and ECMWF-Integrated Forecasting System (ECMWF-IFS) analysis. The configurations and forecasting strategies are the same as those described in Bi23 [4]. In this study, Pangu-Weather was driven by the ERA5 and ECMWF-IFS, respectively, and were denoted as Pangu-ERA5 and Pangu-ECMWF, respectively. It is important to note that due to the delayed availability of the ERA5 product, Pangu-ERA5 is not an operational forecast result, unlike the other forecast data employed in this study [4]. The Pangu-Weather model forecast results provide 6-to-168-hour prediction outputs with a 6-hour temporal resolution.

*4. Human forecast data*

Predictions issued by three meteorological agencies were evaluated. These predictions were made by human forecasters by taking into account their experience and the NWP model simulation results, and are the officially issued forecasts. The three forecast agencies considered in this study include the Japan Meteorological Agency (JMA), China Meteorological Administration (CMA), and Joint Typhoon Warning Center (JTWC). JMA is the regional center designated by the WMO that is responsible for TC warnings and advisories in the WNP. CMA is selected as a representative agency of the WMO members in the WNP, as it issues TC warnings and advisories not only for its land areas and coastal waters, but also for the open seas to meet the increasing demand for disaster prevention and mitigation in various disciplines. While JTWC is not directly associated with the WMO, it is the third agency selected in this study because its warning products are routinely available through both the Global Telecommunication System (GTS) of the WMO and the JTWC's website. The forecast track data of the three agencies can be downloaded from their official websites,

http://www.jma.go.jp/jma/jma-eng/jma-center/rsmc-hp-pub-eg/RSMC_HP.htm, http://typhoon.nmc.cn/web.html, and http://www.metoc.navy.mil/jtwc/jtwc.html, respectively. The TC track forecast data of JMA, CMA, and JTWC used in this study are from the in-house archives of CMA, which were obtained routinely from the WMO GTS in real time. The human forecast results provide 6-to-120-hour prediction outputs.

**Table S1.** Details of the forecast data employed in this commentary.

| Category | Abbreviation | Full name or short description | Source |
|---|---|---|---|
| **NWP** | ECMWF-IFS | Integrated Forecasting System of ECMWF | ECMWF |
|  | NCEP-GFS | Global Forecast System of NCEP | NCEP |
|  | CMA-TRAMS-125L | CMA Regional Typhoon forecasting Model | CMA |
| **AIWP** | Pangu-ECMWF | Pangu-Weather was driven by the ECMWF | Pangu |
|  | Pangu-ERA5 | Pangu-Weather was driven by the ERA5 | Pangu |
|  | Pangu-NCEP | Pangu-Weather was driven by the NCEP | Pangu |
| **Human forecasters** | CMA | Official forecast issued by China Meteorological Administration | CMA |
|  | JMA | Official forecast issued by Japan Meteorological Agency | JMA |
|  | JTWC | Official forecast issued by Joint Typhoon Warning Center | JTWC |

**Text S2. Methods**

*1. Track forecast error*

The track forecast errors are, as done in most studies, estimated by calculating the great circle arc length between the forecast results and the observed TC position. The predicted TC positions were estimated by locating the longitudes and latitudes with the lowest 850-hPa geopotential. We ensure that the evaluation sample sizes of Pangu-ECMWF, Pangu-NCEP, Pangu-ERA5, ECMWF-IFS, NCEP-GFS, and CMA-TRAMS-L125 are consistent for each forecast lead time. However, due to the large amount of missing data of human forecasting results, the track errors of JTWC, JMA, and CMA are calculated based on all samples, which may not be consistent with those of AIWP and NWP models.

*2. Relative forecasting skill*

To compare the forecasting skill among various approaches for different forecast lead times, the relative forecast skills are derived and visualized in Fig. 1f. For each forecast lead time, the relative forecasting skill $RS$ is calculated by the following equations (Eqs. 1–3):

$$RS(i) = -1 * \frac{Err(i) - Avg}{Std} \qquad (1)$$

$$Avg = \frac{\sum_i Err(i)}{N} \qquad (2)$$

$$Std = \sqrt{\frac{\sum_i (Err(i) - Avg)^2}{N}} \qquad (3)$$

where $Err(i)$ denotes the track error of forecast approach $i$, $Avg$ denotes the average track error of all approaches, and $Std$ denotes the standard deviation of the track error of all approaches. A positive $RS$ denotes above-average forecast skills and larger positive values indicate better forecasting skills, and vice versa. In Fig. 1f, only evaluation results of ECMWF-IFS, Pangu-ECMWF, CMA-TRAMS-L125, and human forecasters (average of JTWC, JMA, and CMA) are included in the relative skill calculations.

*3. Evaluation of atmospheric variables*

When evaluating the predicted fields of AIWP and NWP models, we apply the anomaly-based synoptic analysis approach. The anomaly-based synoptic analysis approach provides an effective way to compare the magnitudes and positions of anomalous atmospheric systems in both observation and forecast outputs [5]. To achieve this, we first derive the hourly climatology of each variable based on ERA5 reanalysis data from 1981–2010. Then, the anomaly field of a variable is obtained by subtracting the climatology from the forecast output data. More details about the procedure and physics behind the approach are referred to [5].

**Text S3. Relative performances of Pangu-Weather, NWP models, and human forecasters on predicting Typhoon Khanun's trajectory**

Considering the leading position of ECMWF-IFS and the fact that the Pangu-Weather model was trained based on the ECMWF NWP systems [6], we compare the performances of Pangu-ECMWF with ECMWF-IFS and CMA-TRAMS-L125, as well as the human forecasting results issued by three different centers.

When examining the track error for a 24-48 hour forecast lead time, Pangu-ECMWF demonstrates prediction performance that is as good as ECMWF-IFS and CMA-TRAMS-L125. The average track error of Pangu-ECMWF (58.0 km) is close to that of the NWP models (50.7–55.1 km), and is approximately 27% smaller than those of human forecasters (76.8–82.5 km). However, the gap between Pangu-ECMWF (188.7 km) and the NWP models (123.4–174.7 km) increases to approximately 20% for the 72–96-hour track forecasts, and that between Pangu-ECMWF and human forecasters (207.0–213.4 km) reduces to 10%. This suggests that the Pangu-Weather model experiences a steeper decline in prediction skill as the forecast lead time increases, compared to both human forecasters and NWP models.

Besides, the Pangu-Weather model demonstrates a lower average error (181.4 km) than human forecasters (197.0 km), but it exhibits relatively poorer prediction skills than both the global (151.9 km) and regional (115.2 km) NWP models (Fig. 1e), for 24-to-120-hour track forecast of Khanun. Consistent results are obtained for the prediction uncertainty, highlighting the advantages of the AIWP model over human forecasters but not the NWP models. In addition, the relative performances of AIWP and NWP models vary across different forecast lead times. Pangu-Weather experiences a steeper decline in prediction skill as the forecast lead time increases, compared to both human forecasters and NWP models. The difference between Pangu-ECMWF (379.8 km) and human forecasters (384.7–443.0 km) becomes indistinguishable for a 5-day forecast lead time (Fig. 1e).

Furthermore, for a 120-hour forecast lead time, the difference between Pangu-ECMWF (379.8 km) and human forecasters (384.7–443.0 km) becomes indistinguishable, while ECMWF-IFS (318.2 km) and CMA-TRAMS-L125 (227.9 km) exhibit great advantages in predicting Typhoon Khanun's trajectory (Fig. 1e). These results suggest that although the Pangu-Weather model demonstrates good prediction skills in global-scale forecasts [4], it only catches the performance of NWP models for short-term forecasts of 1-2 days in extreme cases like Typhoon Khanun. For forecast lead times of 3 days or longer, the prediction skills of the Pangu-Weather model decline much more rapidly than those of other forecasting approaches, as indicated by the relative forecasting skill of each prediction approach (Fig. 1f).

**Text S4. Factors leading to differences in prediction performance between Pangu-ECMWF and ECMWF-IFS**

Here, we define 2 groups of TC cases, according to the prediction performance of the ECMWF-IFS and Pangu-ECMWF. (1) The first group includes TCs that are better predicted by the Pangu-ECMWF (hereafter, Group 1); (2) the second group included TCs that are better predicted by the ECMWF-IFS (hereafter, Group 2). Among the 104 TC cases from 2020–2024, 17 cases belong to Group 1. The mean distance error of Pangu-ECMWF ranges from 51.5–230.1 km for 6–120 forecast hours, which are on average 36.6% smaller than those of ECMWF-IFS (ranging from 51.3–363.3 km) (Fig. S5). Meanwhile, 11 cases belong to Group 2. The mean distance error of Pangu-ECMWF ranges from 37.8–372.2 km for 6–120 forecast hours, which are on average 37.9% greater than those of ECMWF-IFS (ranging from 30.3–279.4 km) (Fig. S6). It is important to note that Group 2 TCs are predominantly featured with anomalous trajectories, mostly sudden-turning TCs.

We noticed that Group 1 TCs predominantly exhibit usual moving trajectories, following the large-scale climatological steering flow of the western Pacific subtropical high (WPSH). This is evident in the composite analyses on the background WPSH metrics. In Group1, the flow is mainly steered by southeasterly winds (i.e., negative zonal wind, positive meridional wind in Fig. 2g), and the WPSH has a larger area (Fig. 2h). In this case, TCs mainly move under a relatively strong large-scale steering flow. In Group2, the guiding airflow is mainly southwest winds, and the WPSH has a smaller area, indicating that the steering flow of the environmental field is relatively weak.

Under the influence of a stronger WPSH, TCs' moving trajectories are more likely to follow the background steering flow. On the other hand, under the background of a weaker WPSH, TCs' trajectories are often result of the interaction between the background steering flow and the TC itself, so the prediction accuracies of the fine structural characteristics of the TC core region becomes important. Since AIWP models, such as Pangu-Weather in this case, are more capable of capturing the evolution patterns of large-scale circulations (e.g., WPSH), they tend to give better prediction results for TCs when the WPSH is strong, i.e., Group 1 TCs. On the contrary, AIWP models are not good at "simulating" TC structures and fail to reproduce realistic physics of TC systems. Thus, AIWP models tend to give worse prediction results for the fine structural characteristics of the TC core region when the WPSH is relatively weak, i.e., Group 2 TCs.

Further, we find that similar discrepancies are also observed between sudden-turning TCs and looping TCs. Namely, sudden-turning TCs often occur when the WPSH is relatively weak (Fig. 2e). It is the relatively weak WPSH that leads to sudden-turning TCs more easily affected by the fine structural characteristics of the TC core region, hence resulting in unusual sudden changes in the TC moving direction. As explained above, since AIWP models are not good at "simulating" these finer-scale systems near the TC eyes, the Pangu-Weather model underperforms ECMWF-IFS in predicting these sudden-turning TCs. In contrast, looping TCs are influenced by a comparatively stronger WPSH (Fig. 2f). Thus, the Pangu-Weather model gives better prediction results for these looping TCs, even though they are also considered unusual TC tracks.

**Text S5. Physical relationships among variables in AIWP and NWP predictions**

A crucial reason that makes NWP models being used as the most reliable weather forecast technique for decades is its framework of simulating atmospheric phenomena by solving a set of governing partial differential equations. This framework ensures that the simulation results follow the fundamental laws of physics and atmospheric dynamics, which helps forecasters interpret and analyze model outputs based on their meteorological knowledge. However, Pangu-Weather and AIWP models, in contrast, are trained solely on historical reanalysis data. Current AIWP models do not apply any constraints that ensure the models' ability to 'learn' these physical laws from the input training set. Thus, it is important to know whether Pangu-Weather is able to capture fundamental well-known physical relationships among atmospheric variables, as NWP models do.

First, we examined the TC warm core structure predicted by the Pangu-Weather model. The warm core of a TC is a result of the latent heat release caused by the updraft and condensation of moist air. As shown in Fig. S7, both CMA-TRAMS and ECMWF-IFS, the two NWP models, show a clear warm core structure for TC Khanun, which is centered at approximately 250 hPa. Pangu-ECMWF also successfully reproduced Khanun's warm core, despite its a lower altitude and weaker intensity (Fig. S7d). In addition, the Pangu-Weather model is also able to capture the quasi-hydrostatic relationship between geopotential and temperature. Namely, the TC warm core is associated with a negative geopotential height anomaly center at the bottom [7]. These suggest that the data-driven Pangu-Weather model successfully 'learned' some physical nature of the atmosphere from the input ERA5 training dataset.

Next, we compare the predictions of the geostrophic relationship between geopotential height and horizontal wind between NWP and AIWP models. Geostrophic balance refers to the state of equilibrium between pressure gradient force and Coriolis force, which determines the wind speed and wind direction of mid-latitude large-scale circulation. In real-world situations, the horizontal wind field can be decomposed into the geostrophic component and the ageostrophic component. As shown in Fig. S8, the two NWP models (CMA-TRAMS and ECMWF-IFS) are able to reproduce the observed magnitude of both the geostrophic and ageostrophic components at different isobaric levels. Moreover, the prediction skill is not dependent on the forecast lead time. However, Pangu-ECMWF significantly underestimates the intensity of both the geostrophic and ageostrophic components of wind fields, as well as the ratio between the two components. This suggests that Pangu-Weather has limited ability to reproduce the relationship between geopotential and horizontal wind field in this case.

In addition, we find that the Pangu-Weather model fails to predict the strong vertical wind of TC Khanun although it is able to capture the low-pressure system of Khanun. Khanun, as well as other TCs, are characterized by low pressure and strong updrafts near the TC center, as shown in Fig. S9. It is interesting to note the weak vertical velocity predicted by Pangu-ECMWF (Fig. S9d), even though the low pressure is clearly shown in the forecast output. This is likely because vertical velocity is not included as one of the variables and is not taken into account in the cost function during the training process of Pangu-Weather [4]. This suggests that the data-driven AIWP model does not necessarily capture the physical laws in its prediction results, and further implies the necessity of incorporating physical constraints in the AIWP model.

**Supplementary figures**

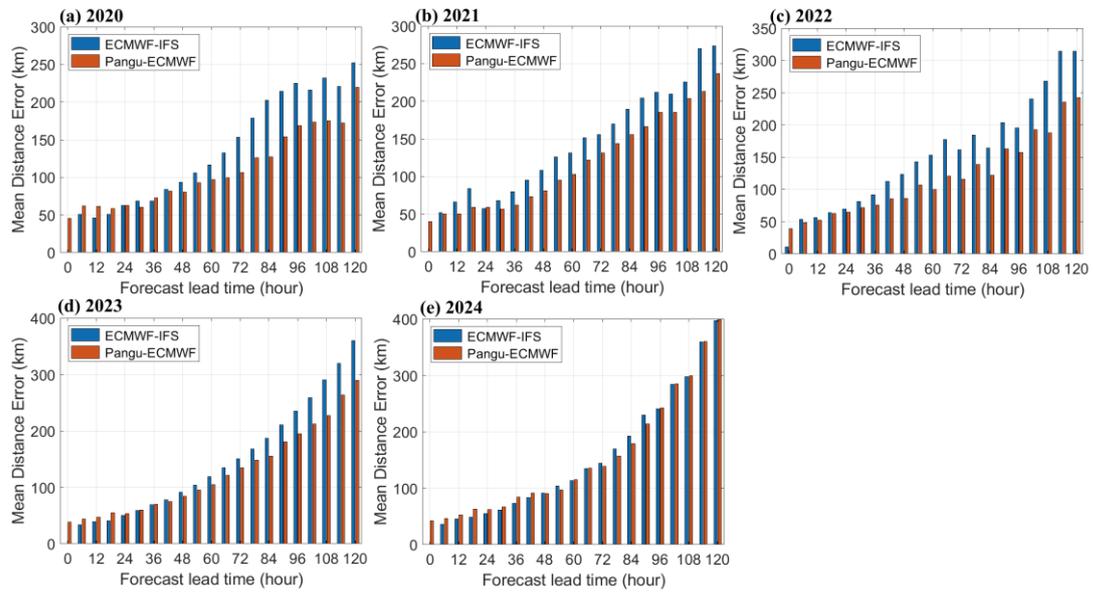

**Fig. S1.** Overall performance of Pangu-ECMWF and ECMWF-IFS in predicting TC trajectories in (a) 2020, (b) 2021, (c) 2022, (d) 2023, and (e) 2024.

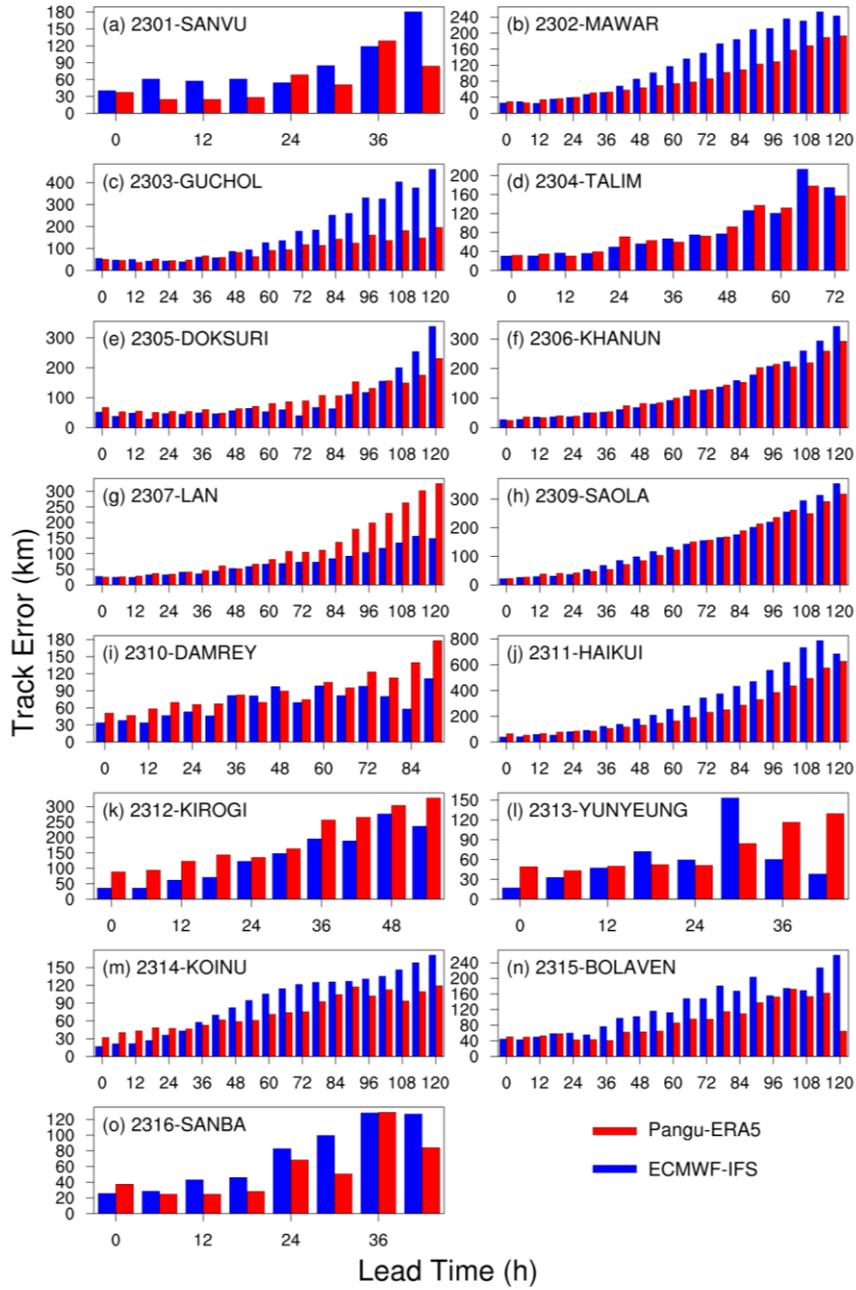

**Fig. S2.** Overall performance of Pangu-ERA5 and ECMWF-IFS in predicting TC trajectories in 2023. (a–o) Average forecast errors of the TC tracks (unit: km) obtained by Pangu-ERA5 (red) and ECMWF-IFS (blue) for (a) TC2301, (b) TC2302, (c) TC2303, (d) TC2304, (e) TC2305, (f) TC2306, (g) TC2307, (h) TC2309, (i) TC2310, (j) TC2311, (k) TC2312, (l) TC2313, (m) TC2314, (n) TC2315, and (o) TC2316, respectively.

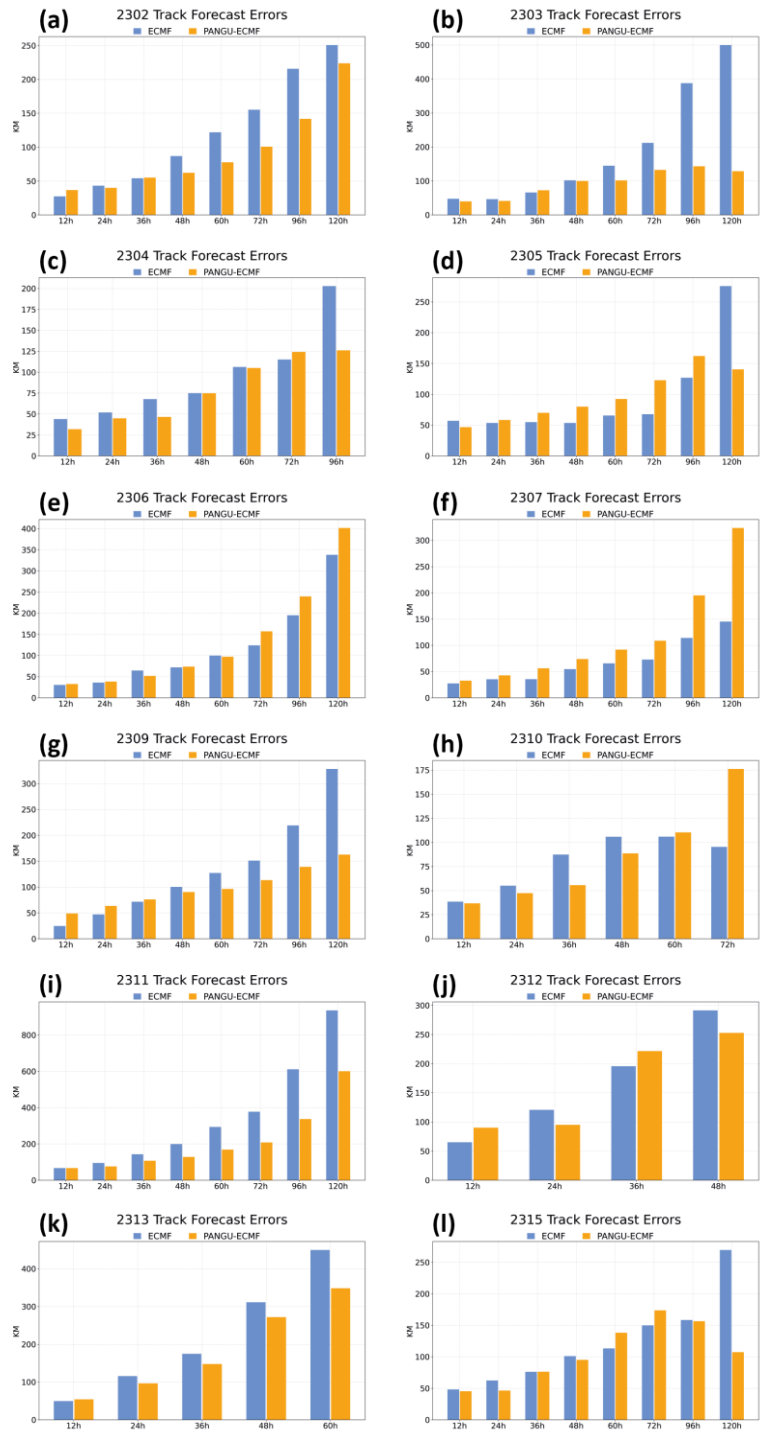

**Fig. S3.** Overall performance of Pangu-ECMWF and ECMWF-IFS for predicting TC trajectories in 2023. (a–o) Forecast errors of TC tracks (unit: km) by Pangu-ECMWF (orange) and ECMWF-IFS (blue) for TC2302 (a), TC2303 (b), TC2304 (c), TC2305 (d), TC2306 (e), TC2307 (f), TC2309 (g), TC2310 (h), TC2311 (i), TC2312 (j), TC2313 (k), and TC2315 (l), respectively.

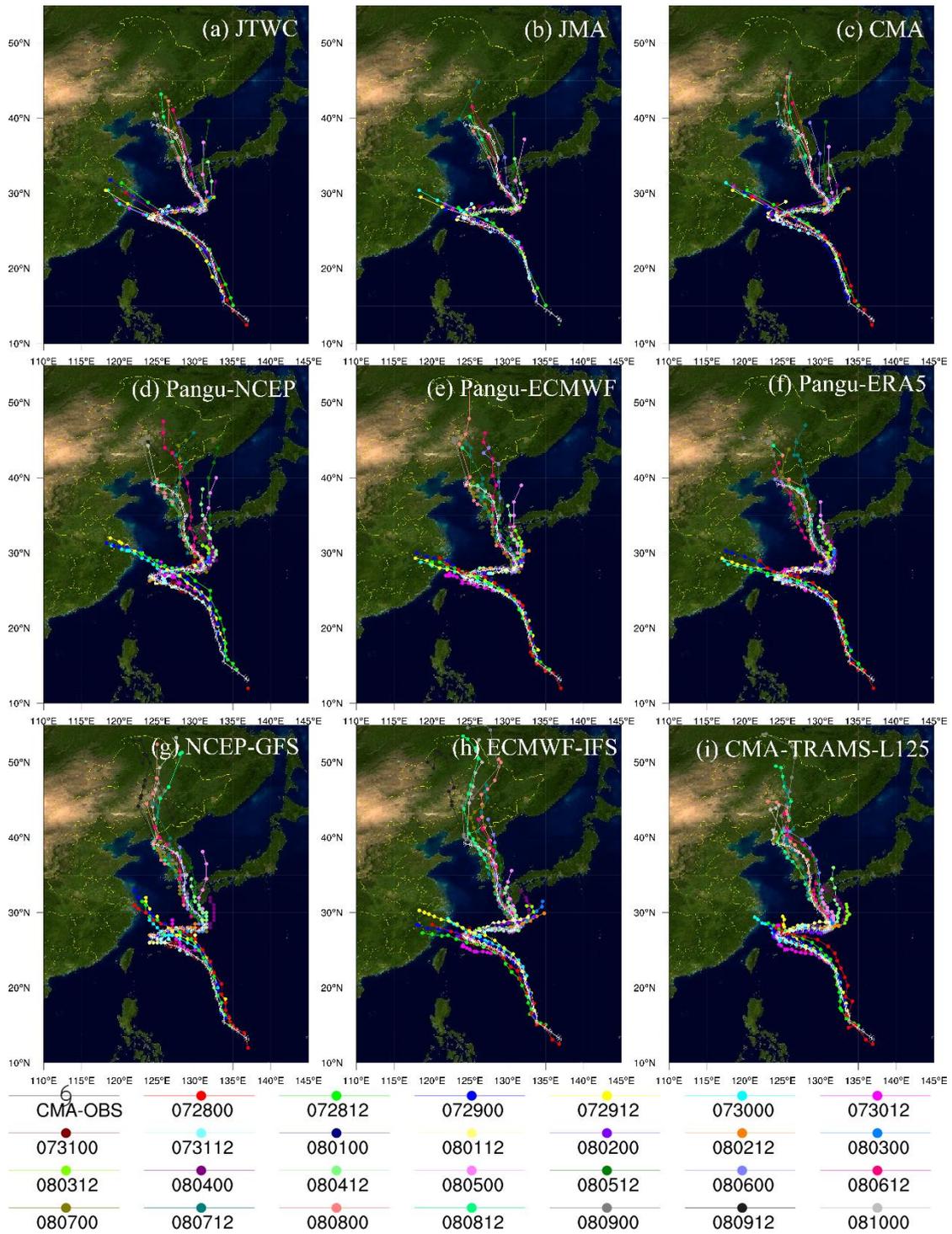

**Fig. S4.** Predicted paths of Typhoon Khanun's by NWP models, AIWP models, and human forecasters. (a–i) 120-hour prediction of Typhoon Khanun's track by (a) JTWC, (b) JMA, (c) CMA, (d) Pangu-NCEP, (e) Pangu-ECMWF, (f) Pangu-ERA5, (g) NCEP-GFS, (h) ECMWF-IFS, and (i) CMA-TRAMS-L125, respectively.

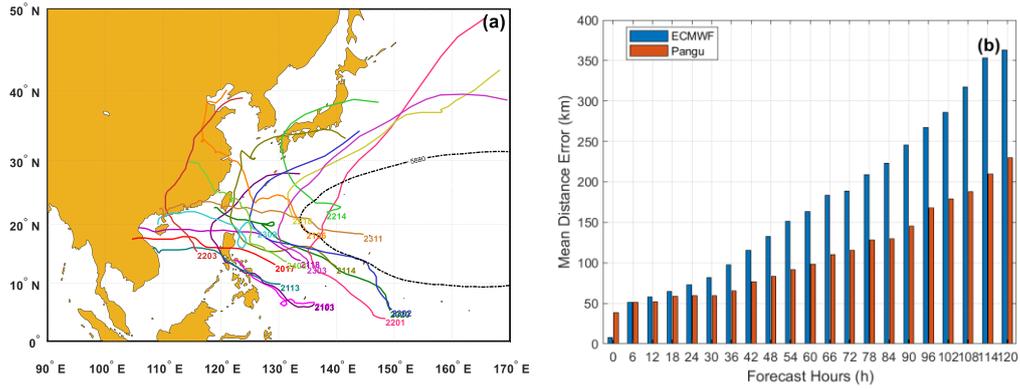

**Figure S5.** (a) Trajectories of Group 1 TCs, which are better predicted by the Pangu-ECMWF, from 2020 to 2024. The black dashed line denotes the averaged WPSH (5880 gpm contour) during the lifetime of all Group 1 TCs. (b) The average error (unit: km) of TC tracks predicted by the ECMWF-IFS (blue) and Pangu-ECMWF (red).

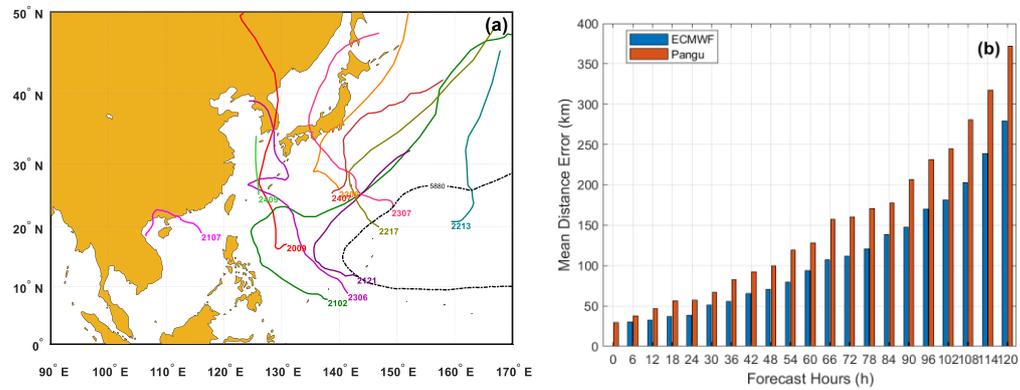

**Figure S6.** (a) Trajectories of Group 2 TCs, which are better predicted by the ECMWF-IFS, from 2020 to 2024. The black dashed line denotes the averaged WPSH (5880 gpm contour) during the lifetime of all Group 2 TCs. (b) The average error (unit: km) of TC tracks predicted by the ECMWF-IFS (blue) and Pangu-ECMWF (red).

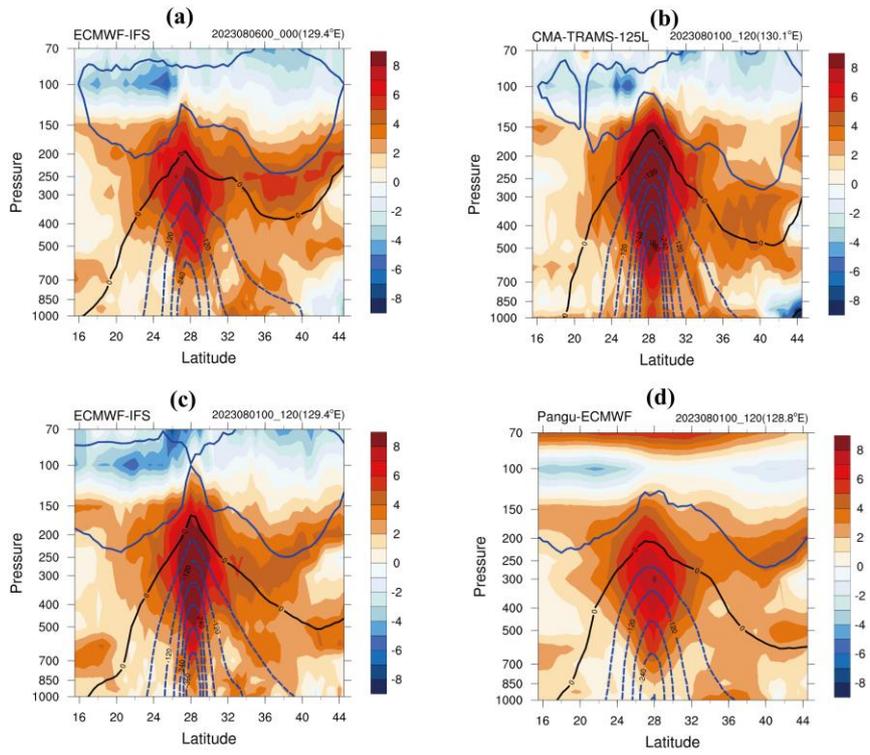

**Fig. S7.** Vertical profile of temperature (shading, unit: K) and geopotential height (contour, unit: gpm) of (a) the ECMWF-IFS analysis, and those predicted by (b) CMA-TRAMS, (c) ECMWF-IFS, and (d) Pangu-ECMWF at forecast lead times of t+120 h, initiated at 0000 UTC 1st August 2023.

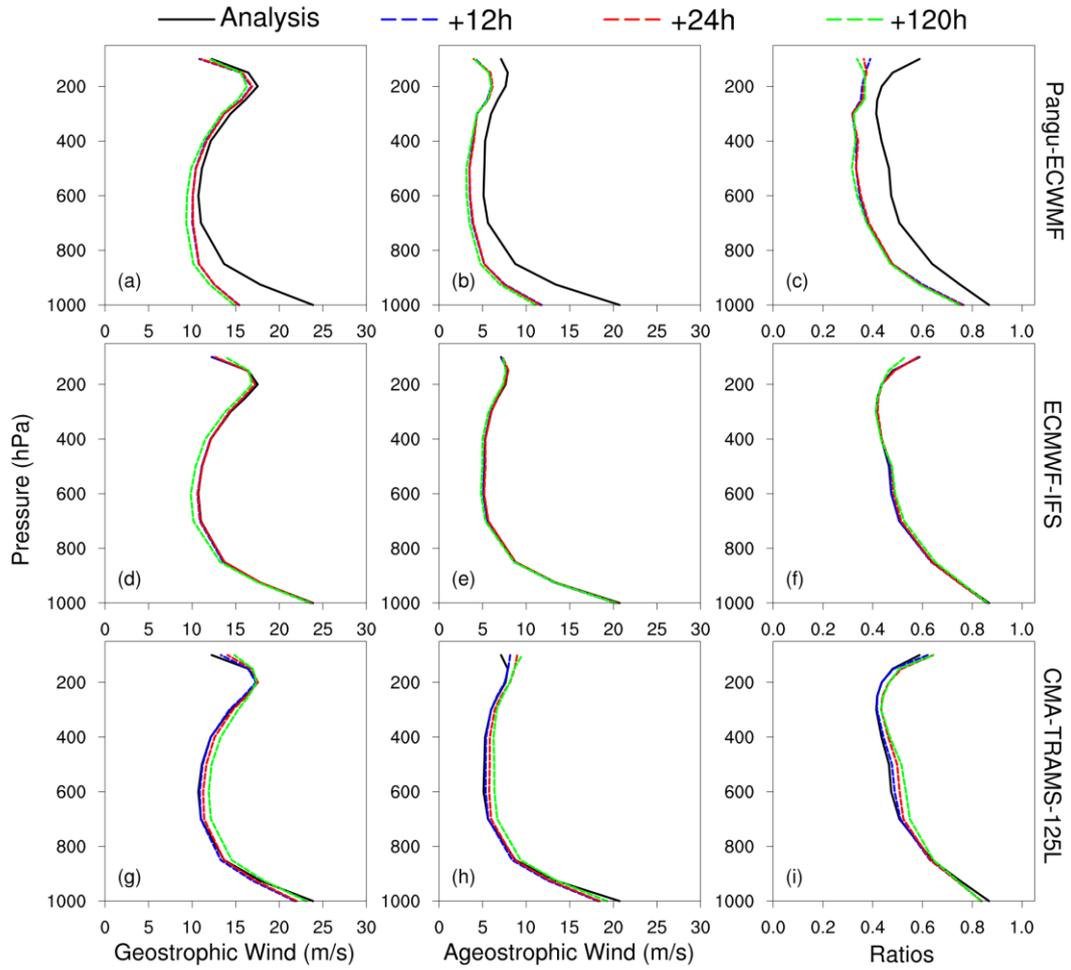

**Fig. S8.** Geostrophic balance predictions by AIWP and NWP models. (a–i) Vertical profiles of the intensity of geostrophic wind (first column), ageostrophic wind (second column) and ratio of the intensity of ageostrophic over geostrophic wind (third column) over the WNP (20°N ≤ Latitude ≤ 50°N & 70°E ≤ Longitude ≤ 160°E) predicted by (a–c) Pangu-ECMWF, (d–f) ECMWF-IFS, and (g–i) CMA-TRAMS at forecast lead times of t+12h (red dashed lines), t+24h (blue dashed lines) and t+120h (green dashed lines), respectively. The black solid lines denote the observed values, obtained from the EMCWF-IFS analysis. The results are averaged over the period from 2023-07-28 to 2023-08-10. The results show that the Pangu-Weather model has the worst ability to predict the accurate relationships between horizontal wind and geopotential fields.

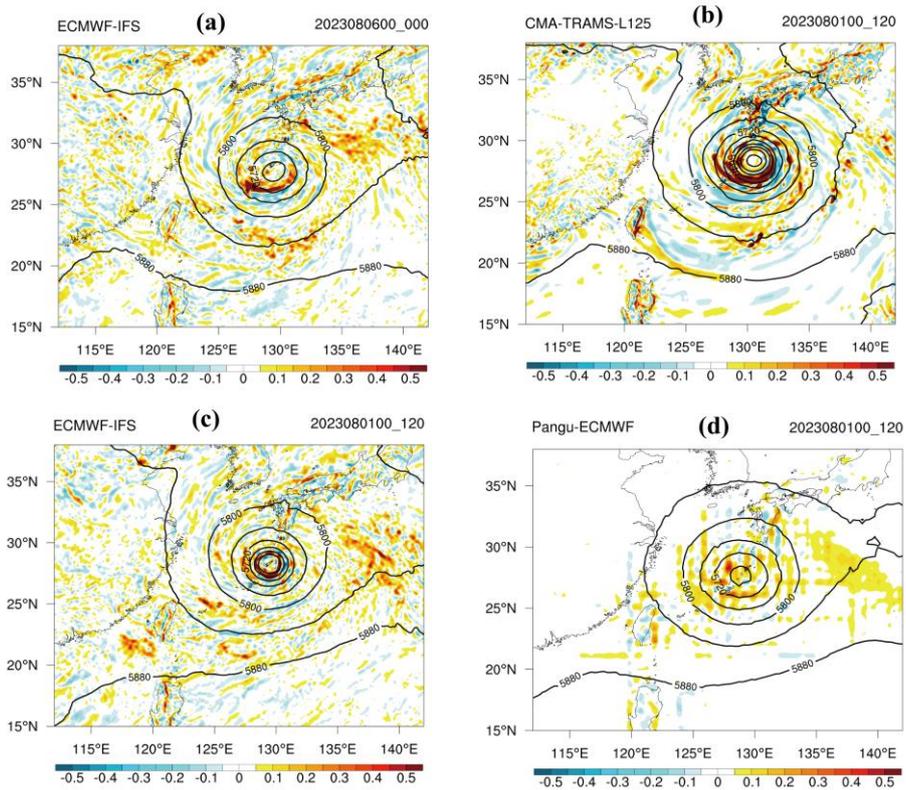

**Fig. S9.** 500-hPa vertical velocity (shading, unit: m/s, derived from the continuity relation) and geopotential height (contour, unit: gpm) of (a) the ECMWF-IFS analysis, and those predicted by (b) CMA-TRAMS, (c) ECMWF-IFS, and (d) Pangu-ECMWF at forecast lead times of t+120 h, initiated at 0000 UTC 1st August 2023.